\documentstyle[12pt]{article}
\input{epsf}

\begin{document}
\begin{titlepage}
\vspace*{-2.2cm}
\begin{center}
\LARGE
   {Internal structure of non-Abelian black holes
    and nature of singularity %
   \footnote{Talk given at the international Workshop
    on The Internal Structure of Black Holes and Spacetime Singularities,
    Haifa, Israel, June 29~-- July 3, 1997.}} \\
\normalsize
\vspace{5mm}
\large{D.~V.~Gal'tsov} \\
\normalsize
       Department of Theoretical Physics, \\
       Moscow State University, 119899 Moscow, Russia \\
\vspace{3mm}
\large{E.~E.~Donets} \\
\normalsize
       Laboratory of High Energies, JINR, 141980 Dubna, Russia \\
\vspace{3mm}
\large{M.~Yu.~Zotov} \\
\normalsize
       Skobeltsyn Institute of Nuclear Physics, \\
       Moscow State University, 119899 Moscow, Russia \\
\end{center}
\vspace{3mm}
\begin{abstract}
 Recent results concerning the internal 
 structure of static spherically-symmetric non-Abelian black holes 
 in the Einstein--Yang--Mills (EYM) theory and its generalizations including 
 scalar fields are reviewed and discussed with an emphasis on the
 problem of a generic singularity in black holes. It is argued that
 in the theories admitting a violation of the naive no-hair conjecture
 the structure of singularity is essentially affected by the ``hair roots''.
 This invalidates an image of a non-Abelian black hole as a Schwarzschild 
 black hole sitting inside the soliton. We give an analytic description of 
 the generic oscillatory approach to the singularity in the pure
 EYM theory in terms of a divergent discrete sequence and show that 
 the mass function is exponentially growing ``in average''. The second type 
 of a generic approach to the singularity in hairy black holes is
 a ``power-law mass inflation'' which is realized in the theories
 including scalar fields. Both singularities are spacelike and no
 Cauchy horizons are met in the full interior region in conformity with
 the Strong Cosmic Censorship conjecture. Black holes violating
 this conjecture exist only for certain discrete values of the event
 horizon radius thus forming a subset of zero measure.

\end{abstract}
\end{titlepage}

\section{Introduction}
\renewcommand{\theequation}{1.\arabic{equation}}

The no-hair conjecture \cite{rw} has played an important
role in understanding the nature of black holes. As a result
of the early investigation of various field theories coupled to gravity 
it was generally believed that the only fields which may extend 
outside the event horizon (apart from external ones not related
to the black hole itself) are those associated with the conserved or 
topological charges carried by the hole.

Later a notable counter example to the naive no-hair conjecture      
was found in the framework of the Einstein--Yang--Mills theory.
Although no classical glueballs may exist in the flat spacetime
because of purely repulsive nature of the constituent vector fields,
such particle-like objects become possible when gravitational attraction 
is taken into account (Bartnik--McKinnon (BK) solutions \cite{bk}). 
It was shown  \cite{sph} that  the lower mass BK particle
topologically is similar to the sphaleron of the Weinberg--Salam theory, 
with a substantial difference, however, due to existence in the BK case
of the gravitational negative mode. Historically, just 
the gravitational instability of the BK solutions was discovered first 
\cite{str} and  later invoked as an argument favoring
the sphaleron interpretation of the BK solutions \cite{sw}. More detailed 
analysis showed, however, that the {\it gravitational\/} (even parity)
negative modes of the BK particles had nothing to do with an expected 
sphaleron picture. Genuine sphaleronic features of BK solutions are 
manifest via the {\it odd-parity\/} negative modes along the vacuum-to-vacuum 
direction in the functional space \cite{snm}. 
Physically the lower mass BK solution may be seen as interpolating
between the neighboring topologically distinct YM vacua in the EYM theory. 
It carries neither electric nor magnetic charge, while the Chern--Simons 
number 1/2 is naturally associated with it \cite{sph,mw}.
 
It was soon realized that the EYM theory also admits the black hole 
counterparts to the BK solutions \cite{vgkb,vg}.
The $SU(2)$ EYM black holes form a two-parameter family
labeled by a continuously varying radius of the event horizon $r_h$
and the number $n$ of oscillations of the YM field in the
strip $[-1, 1]$ bounded by the neighboring YM vacua values. 
Since no asymptotic charges are present, these black holes
apparently violate the no-hair principle. Other solutions
sharing the same property
were found in the non-Abelian gauge theories including scalar fields 
\cite{dil,lnw,gmn} as well as in the gravity coupled Skyrme model.
Meanwhile, a common feeling reasonably persisted  that this kind of hair
should be distinguished from the ``allowed''  hair associated with the 
global charges. It was suggested \cite{tm} to interpret a non-Abelian
field structure outside the event horizon as a ``wig'' 
rather than a genuine hair. Indeed, the EYM 
black hole can lose its YM hair as a result of the gravitational instability
leaving a bare Schwarzschild black hole. As far as the black holes 
inside the magnetic monopoles are concerned, they were often thought of 
(at least for the horizon radius much smaller than the monopole size) 
as tiny Schwarzschild black holes 
sitting inside the solitons \cite{dm}. If so, the inside singularity 
should remain unaffected by the wig. 

However, little was known until recently about the realistic 
internal structure of non-Abelian
black holes (apart from some qualitative discussion in \cite{vg}). Closer
investigation \cite{dgz} has shown that the idea of a Schwarzschild 
black hole sitting inside regular solitons 
is essentially incorrect. The interior structure and the character
of the singularity inside the EYM black hole with an arbitrary 
(continuously varying) radius of the horizon are strikingly different 
from those described by the Schwarzschild geometry.
Both the Schwarzschild and Reissner--Nordstr\"om type singularities
(predicted in \cite{vg}) are encountered indeed, but only for some
discrete values of the horizon radius. This ``second quantization''
in non-Abelian black holes follows mathematically from the same
kind of a non-linear boundary value problem (now in the interior region)
as the ``first quantization'' of the BK particle mass. In this sense
the picture of a Schwarzschild black hole inside the BK particle
covers only the set of solutions of zero measure in the parameter space.
Although it could be anticipated by continuity that the 
non-Abelian hair should penetrate inside the event horizon, 
an essential role of hair in the formation of the singularity seems 
not to have been clearly realized before.
A generic singularity in the spherically symmetric $t$-independent
EYM coupled system turns out to be spacelike and of the oscillatory
nature as was often suggested in view of the Bianchi IX
analysis by Belinskii, Khalatnikov and Lifshitz (BKL) \cite{bkl}. 
In the black hole context, however, the cosmological
counterpart is given by the (non-Bianchi) closed Kantowski--Sachs
cosmology, and the nature of oscillations observed is rather different from
that in the BKL case.

Our results \cite{dgz} concerning the EYM generic black hole 
interior solutions
were confirmed by Breitenlohner et al. \cite{blm} who also attempted
to extend the analysis to monopole black holes in the EYM-Higgs (EYMH)
theory with the triplet Higgs. In the latter case no oscillations were
observed numerically, and the behavior of the mass function near 
the singularity was found to be monotonous. This behavior was 
interpreted in \cite{blm} as exhibiting an exponential
``mass inflation''. Soon after we have found an analytical solution for the 
mass function \cite{gd} in the $SU(2)$ EYMH theories both with the (real)
triplet and (complex) doublet Higgs using a consistent truncation 
of the system of equations near the singularity. Such a truncation is 
possible also in the EYM--dilaton (EYMD) theory \cite{gdz}, 
in both cases the behavior of the
metric near the singularity is dominated by the kinetic (gradient)
contribution of the scalar field. In terms of the radial dependence
of the mass function one finds a {\it power-law\/} divergence towards the 
singularity. Therefore no mass-inflation (in the usual sense \cite{minfl})
is detected inside the EYMH black holes where the singularity is dominated 
by the scalar field. A similar scalar dominated asymptotic behavior
was found earlier in the framework of the Kantowski--Sachs cosmology
\cite{pdm} driven by a non-linear scalar field (in the Abelian model).
In this latter case no black hole counterpart may exist because
of the no-hair theorem for a non-linear one-component
scalar field, but locally the solution is the same. Our formula
for the power-law mass function in the scalar-dominated regime
was reproduced later in \cite{blm2} (see this volume) though without
necessary restrictions on the power index following from the
consistency of the truncation. It is worth noting that
the domains of variation of the power index (depending on 
the horizon radius of the black hole) are different in
the EYMH and EYMD cases, so in the Kantowski--Sachs interpretation
the corresponding singularities look differently. The scalar-dominated
singularity of the ``power-law mass-inflationary'' type inside spherical 
black holes in various field models including scalar fields seems 
to be a rather general phenomenon. This singularity is spacelike 
in conformity with the Strong Cosmic Censorship. Both
oscillatory and the power-law singularities in hairy black holes
therefore support this principle by the genericity argument.

The plan of the talk is as follows. First we discuss the local
solutions near the singularity which may be seen as dressed 
Schwarzschild and Reissner--Nordstr\"om singularities. Then (Sec.~3)
we give an analytic description of the oscillatory approach to the
singularity in the generic EYM black holes. In Sec.~4 it is
shown that no exponential mass inflation develops inside the EYMH and EYMD
black holes where the typical power-law divergence of the mass
function is met. We conclude with a brief discussion
of a new role of the black hole hair which was revealed
in the investigation of black hole interiors.

\section{Hairy Schwarzschild and \protect\\
 Reissner--Nordstr\"om singularities}
\renewcommand{\theequation}{2.\arabic{equation}}
\setcounter{equation}{0} 
We start with the pure $SU(2)$ EYM system
\begin{equation}
                S = \frac{1}{16\pi} \int \left \{-R+ F^2 \right \}
                \sqrt{-g} d^4x \,,
\label{eq1}
\end{equation}
where $F$ is the $SU(2)$ field, assuming the metric to be invariant under
the time translations and the $SO(3)$ spatial rotations:
\begin{equation}
        ds^2 = \frac{\Delta \sigma^2}{r^2} dt^2 - \frac{r^2}{\Delta}dr^2
               -r^2 (d \theta^2 + \sin^2 \theta d \phi^2) \,.
\label{eq2}
\end{equation}
Here the metric functions $\Delta$, $\sigma$ depend only on $r$.
This parameterization of spacetime is suitable until a turning point
of $r$ is reached. It happens that for asymptotically flat solutions 
there are no such points neither in  exterior, nor in interior regions
up to the curvature singularity at $r=0$. Such points exist,
however, for asymptotically non-flat solutions, in which case another
chart should be used \cite{blm}. Here we will not be dealing with
{\it a priori\/} asymptotically non-flat solutions, so the coordinates 
(\ref{eq2}) are good both in the exterior and interior regions.
Of course one should keep in mind that 
in the region where $\Delta<0$ the radial coordinate $r$ is  timelike, 
while $t$ is spacelike, so that $t$-independence means spatial
homogeneity of the spacetime rather than staticity.

As usual, we choose the $t$-independent spherically symmetric magnetic 
ansatz for the YM potential
\begin{equation} \label{eq3}
 A=\left(W(r)-1 \right) \left(T_{\varphi} d \theta-
 T_{\theta} \sin \theta d \varphi \right),
\end{equation}
where $T_{\varphi , \theta}$ are
spherical projections of the $SU(2)$ generators. The value $W=1$ is 
a trivial YM vacuum, while $W=-1$ corresponds to the vacuum of 
the next topological sector (this state is related to the trivial
vacuum by a ``large'' gauge transformation).

 The field equations consist of a coupled system for $W$, $\Delta$
\begin{eqnarray}
     && \Delta U' + \left(1 - \frac{V^2}{r^2} \right)W' = \frac{WV}{r} \, ,
        \label{eq4} \\
     && \left(\frac{\Delta}{r} \right)' + 2 \Delta U^2 = 1 - \frac{V^2}{r^2}
        \, , \label{eq5}
\end{eqnarray}
 where $V=(W^2-1)$, $U=W'/r$, and a decoupled equation for $\sigma$:
\begin{equation} \label{eq6}
 \frac{d}{dr^2} \ln \sigma = U^2 \,.  
 \end{equation}

 These equations admit black hole solutions in the domain
 $r \ge r_h$ for any  radius of the event horizon $r_h$ \cite{vgkb}. 
 The solutions for $W$ outside the horizon lie within the strip
 $-1<W<1$ approaching $\pm 1$ asymptotically. They
 are specified by the number $n$ of nodes of $W$ thus forming
 a discrete set for each $r_h$. 
 Local solutions in the vicinity of the regular event horizon of a given
 radius contain one free parameter $W(r_h)$ \cite{vg}.
 A ``quantization'' of $W(r_h)$ results from an imposition of the
 boundary conditions at infinity.  
 Each asymptotically flat exterior solution starts with
 some uniquely determined value $W^n(r_h)$ at the horizon. All other
 $W(r_h)\neq W^n(r_h)$ generate asymptotically non-flat solutions.
 This quantization does not lead to the quantization of the black hole mass 
 since the radius of the horizon still remains an arbitrary continuously 
 varying parameter.

To start the analysis of the internal structure of non-Abelian
black holes, which is an essentially numerical problem, one
has to explore first local solutions of the field equations
near the singularity. Two such local branches were found
earlier \cite{vg}. The first one is the Schwarzschild--like (S) solution, 
it corresponds to the vacuum value of the YM field $|W(0)|=1$. 
Introducing the mass function $m(r)$, $\Delta=r^2 - 2 m r$, one has
\begin{eqnarray} \label{s}
      &&W=-1+b r^2 +b^2(3-8b)r^5/(30m_0) + O(r^6) \, , \nonumber \\
      &&m=m_0 (1-4b^2r^2+8b^4r^4)+2b^2r^3+O(r^5) \, ,
\end{eqnarray}
 where $m_0$, $b$ are (the only) free parameters. This local 
 branch is non-generic already by counting free parameters (a generic
 solution of the system (\ref{eq4}, \ref{eq5}) should have three 
 free parameters).

 The second is the Reissner--Nordstr\"om  type branch which can be
 found assuming the leading term of $\Delta$ to be a positive constant
 (related to the charge parameter $P^2=\Delta(0)$).
 This  requires $W(0)= W_0 \ne \pm 1$ and gives \cite{vg}
\begin{eqnarray}\label{rn}
      &&W=W_0-W_0 r^2/(2V_0)+c r^3/(2V_0)+O(r^4) \, , \nonumber \\
      &&\Delta=V_0^2  - 2 m_0r + r^2 +2W_0(c+m_0W_0/V_0^2)r^3+O(r^4) \, ,
\end{eqnarray}
 what corresponds to the RN metric of the mass $m_0$ and the (magnetic)
 charge $P^2=V_0^2$, $V_0=V(W_0)$. The expansion contains three free
 parameters $W_0$, $m_0$, $c$ (i.e., is locally generic).
 
These two local solutions may be regarded as describing ``hairy''
Schwarz\-schild and  Reissner--Nordstr\"om singularities. Their
essential distinction from the usual Schwarzschild and  Reissner--Nordstr\"om 
singularities consists in presence of the additional free parameters
($b$ in the first case and $c$ in the second) responsible for
hair degrees of freedom.

 We have also found the third local power series solution \cite{dgz}
 assuming a {\it negative\/} value for $\Delta(0)$ (i.e., {\it imaginary\/} 
 $P$):
\begin{eqnarray} \label{new}
 &&W=W_0 \pm r - W_0 r^2/(2 V_0) +O(r^3) \, , \nonumber \\
 &&\Delta=-V_0^2 \mp 4 W_0 V_0 r+O(r^2) \, , \\
 &&\sigma=\sigma_1 (r^2 \mp 4 W_0 r^3/V_0) + O(r^3) \,.\nonumber
\end{eqnarray}
 Here there is only one free parameter $(W_0)$ for $W$, $\Delta$.
 The corresponding space-time near the singularity
 is conformal to $R^2 \times S^2$: after a time rescaling one obtains
\begin{equation}
 ds^2=r^2(dr^2 -dt^2 - d \theta^2-\sin^2 \theta d \varphi^2) \,.
\end{equation}
 This geometry was encountered in the previous study of black hole interiors
 in the framework of the perturbed Einstein--Maxwell theory \cite{pg,abc} and 
 called homogeneous mass-inflation model (HMI).

 It is easy to realize that neither of these asymptotics may correspond to
 a {\it generic\/} black hole. Imposing boundary conditions in the 
 singularity,  we obtain the same kind of the singular boundary 
 value problem as one encountered in the exterior problem where a similar 
 role is played by the asymptotic flatness condition. 
 This interior boundary value problem leads to the {\it second
 quantization\/} condition, now for the event horizon radius $r_h$. Therefore, 
 the EYM black holes with the S and RN type interiors may constitute 
 only the zero measure set in the whole EYM black hole solution space. 

\begin{figure}
 \centerline{\epsfxsize=13cm \epsfbox{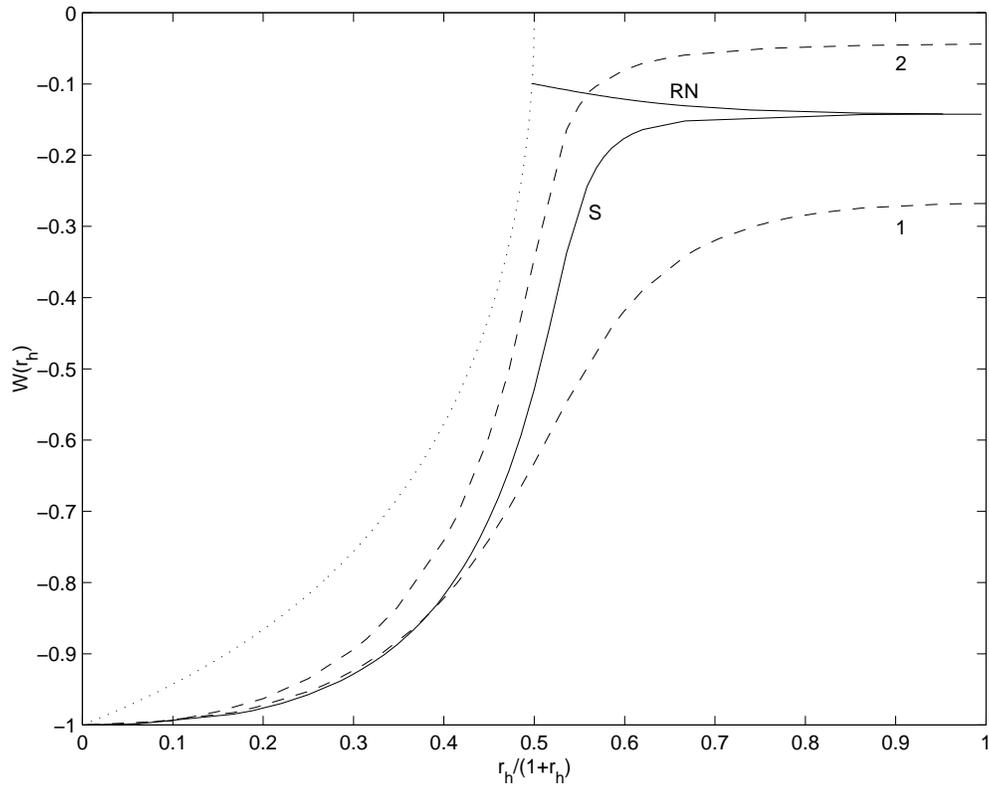}}
 \caption{$W(r_h)$  for the S-- and RN--type interior solutions.
  Dashed lines~--- $W^n(r_h)$ for $n=1, 2$ (higher--$n$ curves lie
  between  the $n=2$ one and the boundary $r_h=1-{W_h}^2$, dotted line).
  Note that S and RN curves $W(r_h)$ do not merge.}
 \label{curves}
\end{figure}

\begin{figure}
 \centerline{\epsfxsize=13cm \epsfbox{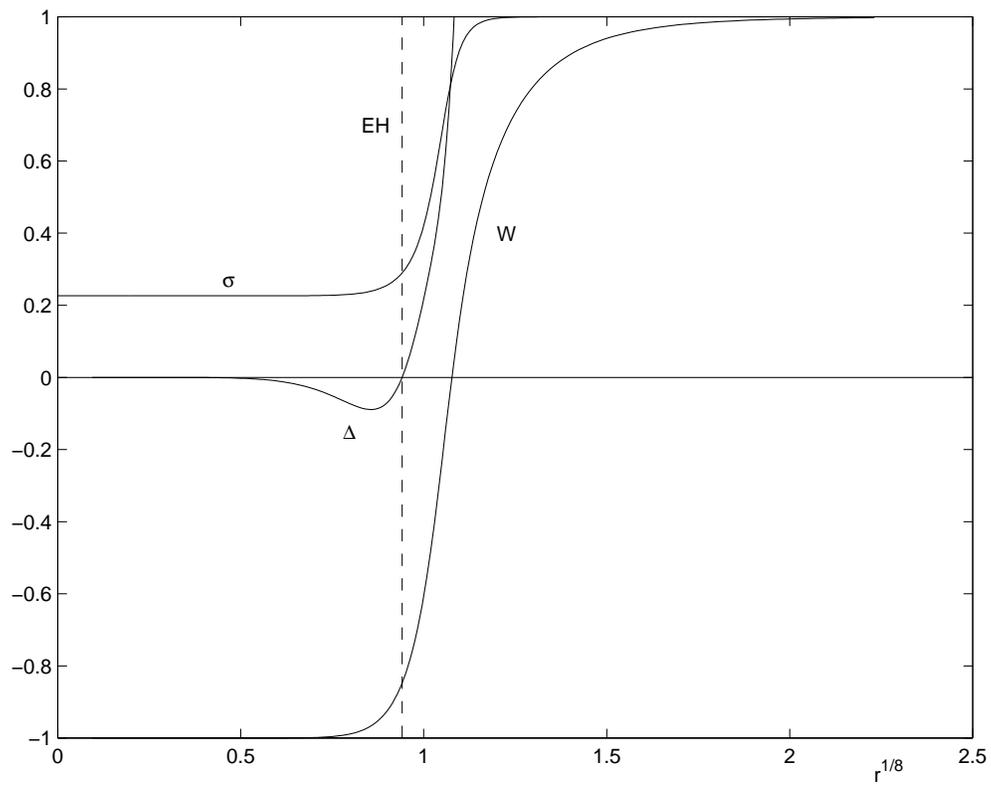}}
 \caption{The $n=1$ EYM black hole (S--type). EH~--- events horizon.}
 \label{schwarz}
\end{figure}

 The system (\ref{eq4}--\ref{eq5}) was integrated numerically
 in the region $0<r<r_h$ using an
 adaptive step size Runge--Kutta method for various $r_h=10^{-8},...,10^{6}$.
 The integration started at the left vicinity of the event horizon $r_h$ 
 where the local power series solution contains one free parameter 
 $W_h = W(r_h)$ satisfying the inequalities  $|W_h|<1$ and $1-W_h^2<r_h$ 
 which are the necessary conditions for the asymptotic flatness \cite{vg}.
 For given $r_h$, the interior solutions meeting the
 expansions (\ref{s}--\ref{new}) may exist only for
 some discrete $W_h$. A numerical strategy used to find
 such $W(r_h)$ consisted in detecting the change of sign of the derivative
 $W'$. In the S-case we found the curve $W(r_h)$ which
 starts at $-1$ as $r_h \rightarrow 0$ and approaches $-0.1424125$ for large
 $r_h$ (Fig.\ \ref{curves}) (without loss of generality we choose
 $W_h<0 $).  Our S-curve intersects the $n=1$ branch of the family 
 of trajectories $W^n(r_h)$ corresponding to the set
 of external asymptotically flat solutions.  Parameters of this black hole 
 are shown in Tab.\ \ref{numres}, its global behavior is depicted
 in Fig.\ \ref{schwarz} (for higher $n$ S-solutions see \cite{blm}).

\begin{table}
 \caption{S-- and RN--type solutions.}
 \begin{center}
  \begin{tabular}{cccc}
  \hline
   & S--type, $n=1$ & RN--type, $n=2$ & RN--type, $n=3$\\
  \hline
   $r_h$     &$0.613861419$&$1.273791$&$1.0318420$\\
   $W(r_h)$  &$-0.8478649145$&$-0.113763994$&$-0.10185163$\\
   $r_{\_}$  & --- &$0.02171654$&$0.08948446$\\
   $W(0)$    &$-1$&$-1.212296124$&$-1.3566052$\\
   $\sigma(0)$&$0.2263801$&$5.991210 \times 10^{-3}$&
   $1.751928 \times 10^{-3}$\\
   Mass      &$0.8807931$&$1.018002$&$1.000277$ \\
  \hline
  \end{tabular}
 \end{center}
 \label{numres}
\end{table}

\begin{figure}
 \centerline{\epsfxsize=13cm \epsfbox{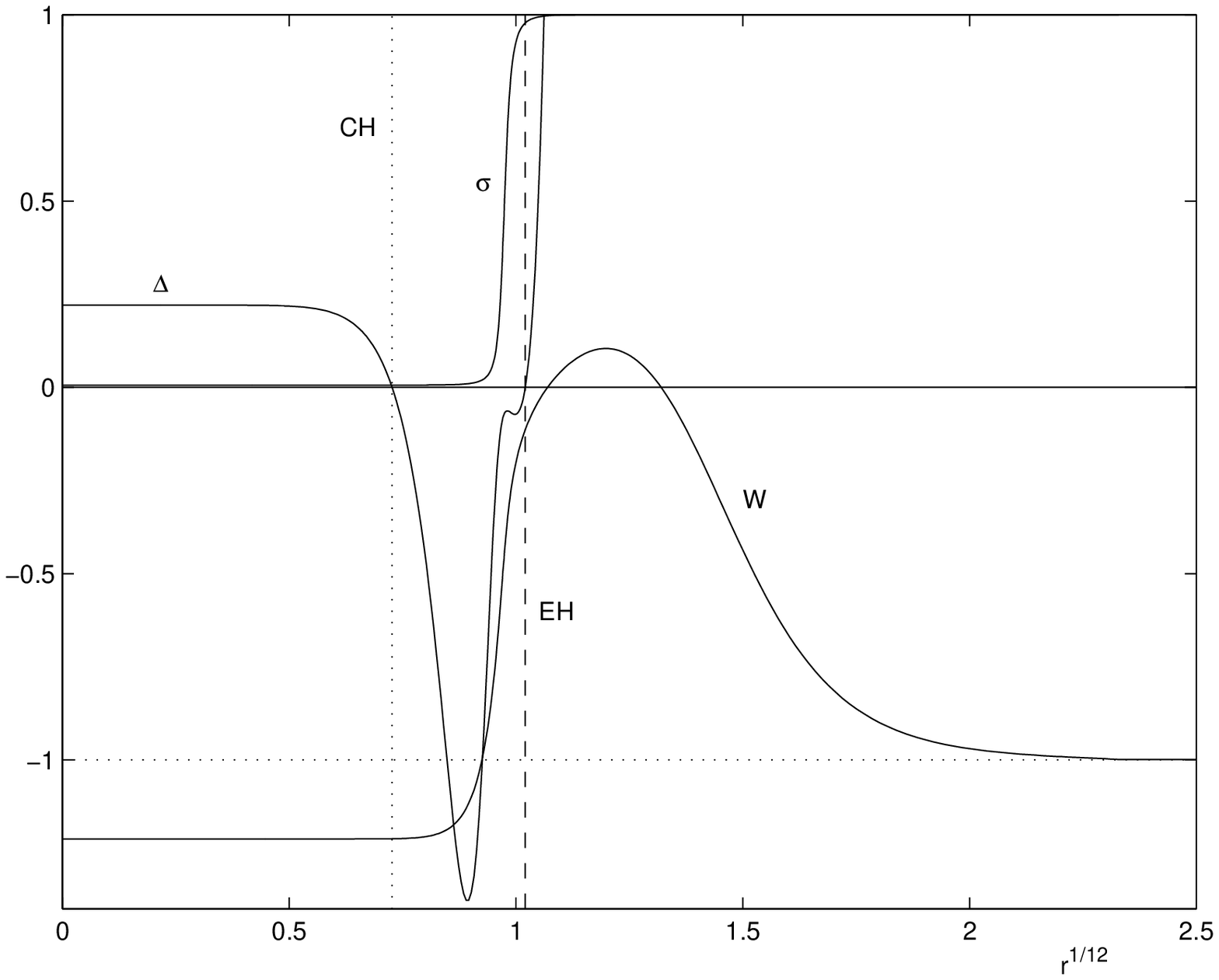}}
 \caption{The $n = 2$ EYM black hole (RN--type). EH~--- events horizon,
          CH~--- Cauchy horizon.}
 \label{RN2}
\end{figure}

 Interior solutions of the RN-type, meeting the expansions
 (\ref{rn}) in the singularity, were found for $r_h >r_h^*=0.990288617$.
 The corresponding curve $W(r_h)$ (also shown in Fig.\ \ref{curves})
 intersects the trajectories $W^n(r_h)$ for all $n \ge 2$.
 These solutions possess an inner Cauchy horizon at some $r_{\_} < r_h$ with
 $|W(r_{\_})|>1$ (Fig.\ \ref{RN2}).

%%%%%%%%%%%%%%%%%%%%%%%%%%%%%%%%%%%%%%%%%%%%%%%%%%%%%%%%%%%%%%%%%%%%%%%%%%%%%%

%%%%%%%%%%%%%%%%%%%%%%%%%%%%%%%%%%%%%%%%%%%%%%%%%%%%%%%%%%%%%%%%%%%%%%%%%%%%%%

 Solutions of the third type (\ref{new}) were studied numerically
 starting from the vicinity of the origin.
 The unique solution has been found for the horizon data subject to the
 necessary  conditions for an asymptotic flatness $|W_h|<1,\; 1-W_h^2<r_h$
 for the upper sign in (\ref{new}) and $W(0)=-0.9330656$, corresponding 
 to $r_h=1.889088$. This solution, however, does not 
 meet any value $W^n(r_h)$ and thus does not represent a black hole. 
 Thus a pure HMI interior can not be attributed to the EYM black holes.
 But it turns out to be an unstable fixed point of an asymptotic 
 (truncated) dynamical system \cite{dgz}. Deviations from this fixed point
 correspond to generic EYM interior solutions sharing the same
 property to possess no Cauchy horizons. 
 
 More general families of internal solutions not restricted by the 
 asymptotical flatness condition were found in \cite{blm} (in particular,
 other internal solutions of the third type). They do not have, however, a
 direct significance for the EYM black holes.

\section{Oscillatory approach to singularity}
\renewcommand{\theequation}{3.\arabic{equation}}
\setcounter{equation}{0} 
 
\begin{figure}
 \centerline{\epsfxsize=13cm \epsfbox{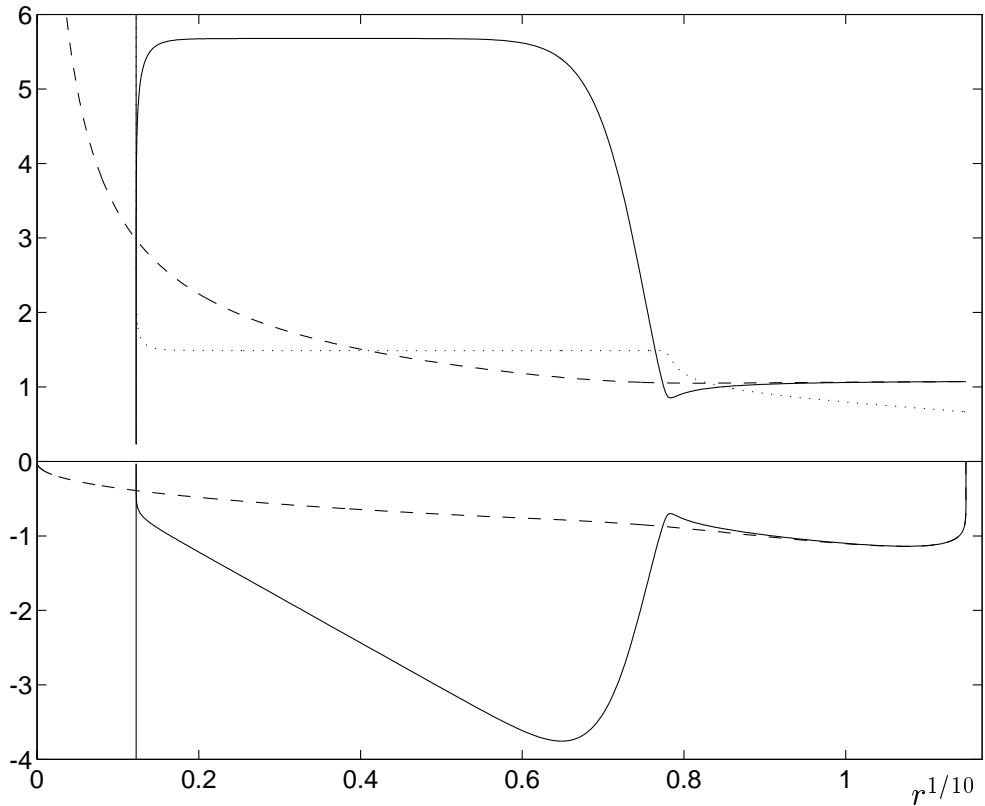}}
 \caption{
    The first oscillation cycle for an EYM solution. 
    The functions $\Delta$ for EYM (solid line) and EYMD (dashed)
    are shown in lower half-plane. In upper half-plane~---
    mass functions $m(r)$ (analogously) and the function $U$ for EYM (dotted). 
    All functions are power rescaled with the power index 1/10.
    Here $r_h = 4$; $W_h = -0.283993$ for EYM,
    $W_h = -0.298357$, $\phi_h = 0.05623$ for EYMD
    (asymptotically flat solutions with one node of $W$.)}
 \label{firstoscln}
\end{figure}

 Since both hairy Schwarzschild and hairy Reissner--Nordstr\"om singularities
 are encountered only for certain discrete values of $r_h$ 
 (and hence the black hole masses), while the external solutions exist
 for continuously varying $r_h$, we must look for an alternative regime
 of approach to the singularity for generic $r_h$. 
 It is generically observed during numerical integration from the events
 horizon towards the origin that sooner or later (depending on $r_h$
 and $W_h$) the metric function $\Delta$ starts to oscillate in the
 negative region with a very fast growing amplitude \cite{dgz}. 
 Because of huge numbers encountered in these oscillations the accuracy
 of the computation can hardly be maintained, so one has to seek 
 appropriate truncations of the system of equations to describe 
 the asymptotic regime analytically. Numerically it is observed that, 
 when oscillation progress, the right hand side of the Eq.\ (\ref{eq4}) 
 becomes small with respect to terms at the left hand side. 
 Neglecting it one obtains the following approximate first
 integral of the system:
\begin{equation}
     Z = \Delta U \sigma/r = {\rm const} \, , \label{eq10}
\end{equation}
 which relates oscillations of the mass function to the evolution of
 $\sigma$. Numerical experiments also show that while the YM function $W$
 remains almost constant up to $r=0$, its derivative is still non-zero and
 is rapidly changing on some very small intervals of $r$. The function $U$
 exhibits a step-like behavior being  constant with high accuracy during 
 almost all the chosen oscillation cycle (Fig.\ \ref{firstoscln}) and then 
 jumping to a greater absolute value corresponding to the next
 cycle. It is clear from (\ref{eq6}) that $\sigma$ is exponentially falling
 down with decreasing $r$ while $U \approx {\rm const}$, 
 whereas in the tiny intervals
 of $U$--jumps $\sigma $ remains almost unchanged.
 So $\sigma$ tends to zero through an infinite sequence of exponential
 falls with increasing powers in the exponentially decreasing intervals. 
 Combining this with (\ref{eq10}) and the above mentioned properties
 of $U$ one can deduce rather detailed description of the metric behavior. 

%%%%%%%%%%%%%%%%%%%%%%%%%%%%%%%%%%%%%%%%%%%%%%%%%%%%%%%%%%%%%%%%%%%%%%%%%%%%%%
%%%%%%%%%%%%%%%%%%%%%%%%%%%%%%%%%%%%%%%%%%%%%%%%%%%%%%%%%%%%%%%%%%%%%%%%%%%%%%

 Let us denote by $r_k$ the value of radial coordinate where $\Delta$ has
 $k$-th local maximum. Soon after passing this point, the function
 $U$ stabilizes at some value $U_k$ approximately equal to the doubled
 value at the point of local maximum (similarly, $U$ increases by about
 a factor of two when approaching the local maximum, whereas $\Delta$
 is almost stationary). Then, according to (\ref{eq6}), $\sigma$ is equal to
\begin{equation}
     \sigma(r) = \sigma(r_k) \exp \left[ U_k^2 (r^2 - r_k^2) \right] \, ,
\end{equation}
 From (\ref{eq10}) one finds that, while $U_k \approx {\rm const}$,
\begin{equation}
     \Delta(r) = \frac{\Delta(r_k)}{r_k} \; r \; 
                 \exp \left[ U_k^2 (r^2_k - r^2) \right] \,. \label{eq11}
\end{equation}
 This function falls down with decreasing $r$ until it reaches a local 
 minimum at 
\begin{equation}
     R_k = \frac{1}{\sqrt{2}\;|U_k|} \approx
            \frac{\sqrt{|\Delta(r_k)|}}{2|V(r_k)|} r_k \,. \label{eq12}
\end{equation}
 In what follows, in view of the observed fact that in the course of 
 oscillations of $\Delta$ the YM function $W$ changes insignificantly, 
 we will put $V={\rm const}$.

 Therefore, the mass function is inflating exponentially while $r$
 decreases from $r_k$ to $R_k$. After passing $R_k$, an exponential in 
 (\ref{eq11}) becomes of the order of unity,  hence $\Delta$ starts to grow
 linearly, and the mass function $m(r)$ stabilizes at the value 
 $M_k = m(R_k)$. Such a behavior holds until the point of local maximum of
 $\Delta/r^2$ is reached; this takes place when $\Delta \approx -V^2$ at
 the point
\begin{equation} 
     r^*_k \approx \frac{V^2}{|\Delta(r_k)|} r_k
                   \exp \left[-(U_k r_k)^2 \right] \,. \label{eq13}
\end{equation} 
 After this a rapid fall of $|\Delta|$ is observed causing a violent rise
 of $|U|$. Then the term $2\Delta U^2$ in the Eq.\ (\ref{eq5}) becomes
 negligible and consequently at this stage
\begin{equation} 
     U \Delta \approx -V^2 U_k \, , \label{eq14}
\end{equation} 
 while $r$ practically stops. This implies that very soon
 $\Delta$ reaches the next local maximum at the point
 $r_{k+1} \approx r_k^*$, while $m(r)$ rapidly falls down to
 $m_{k+1}$. At the point of local maximum of  $\Delta$ 
 one has in the Eq.\ (\ref{eq6}) $|\Delta| \ll V^2$, 
 then in view of the smallness of $r$ we find
\begin{equation} 
     |U(r_k)|  \approx \frac{|V|}{\sqrt{2|\Delta(r_k)|} r_k} \,. \label{eq15}
\end{equation} 

 To obtain the estimates by the order of magnitude we will neglect
 all numerical coefficients elsewhere except for the power indices 
 in exponentials, in particular putting $U(r_k) = U_k$, and omitting
 also (quasi-constant) factors $V$. With this accuracy one obtains from 
 (\ref{eq11})--(\ref{eq15}):
\begin{equation}
     r_{k+1} = M_k^{-1} \, , \quad r_{k+1}^2 = R_k R_{k+1} \, ,\quad
     M_k = \frac{R_k^2}{r_k^3} \exp \left( \frac{r_k^2}{2 R_k^2} \right) \,,
\end{equation}
\begin{equation}
     |\Delta(r_k)| = \left( \frac{R_k}{r_k} \right)^2 \, , \quad
     \frac{r_{k+1}}{r_k} = \frac{r_k^2}{R_k^2} \exp \left[ -\left( 
     \frac{r_k^2}{2 R_k^2} \right) \right] \,.
\end{equation}
 Thus, introducing a variable $x_k=(r_k/R_k)^2 \, (\gg 1)$, we can derive 
 the following recurrent formula
\begin{equation}
     x_{k+1} = x_k^{-3} {\rm e}^{x_k} \, ,
\end{equation}
 which shows that $x_k$ is an exponentially diverging sequence.
 In terms of $x_k$ one has
\begin{equation}
     \frac{r_{k+1}}{r_k} = x_k {\rm e}^{-x_k/2} \, ,
\end{equation}
 this relation can also be understood as a ratio of the neighboring
 oscillation periods since $r_k \gg r_{k+1}$. Values of the function
 $|\Delta|$ at the points $r_k$ rapidly tend to zero:
 \begin{equation}
     |\Delta(r_k)| = x_k^{-1} \, ,
\end{equation}  
 so we deal with an infinite sequence of ``almost'' Cauchy horizons
 as $r \rightarrow 0$. At the same time the values of  $|\Delta|$ at the
 points $R_k$ grow rapidly:
\begin{equation}
     |\Delta(R_k)| = x_k^{-3/2} {\rm e}^{x_k/2} \,,
\end{equation}  
 and the values of the mass function grow correspondingly as
\begin{equation}
     \frac{M_k}{M_{k-1}} = x_k^{-1}{\rm e}^{x_k/2} \,.
\end{equation}  

 While $r$ decreases form $r_k$ to $R_k$, the function $\sigma$ 
 rapidly falls down to the value $\sigma_k = \sigma(R_k)$, which then
 remains unchanged until the point $r_{k+1}$. As the singularity
 is approached, the sequence  $\sigma_k$ decreases according to
\begin{equation}
     \frac{\sigma_{k+1}}{\sigma_k} = {\rm e}^{-x_k/2} \,.
\end{equation}
 Therefore for a generic (continuously varying) $r_h$ one observes 
 a rather unexpected approach to the singularity through an infinite
 sequence of oscillations with an exponentially growing amplitude.

\begin{figure}
 \centerline{\epsfxsize=13cm \epsfbox{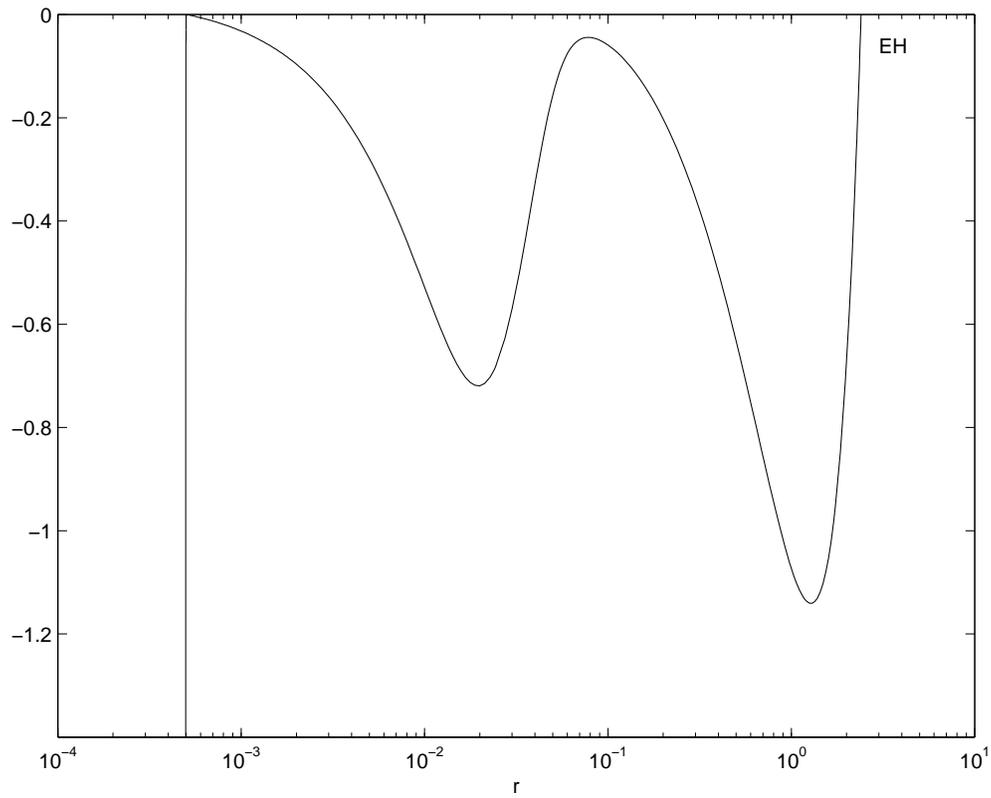}}
 \caption{The beginning of $\Delta$--oscillations  for $n=1$
  EYM BH solution, $r_h=2.4$, $W(r_h)=-0.31652531$. EH~--- events horizon.}
 \label{oscilns}
\end{figure}

\begin{figure}
 \centerline{\epsfxsize=13cm \epsfbox{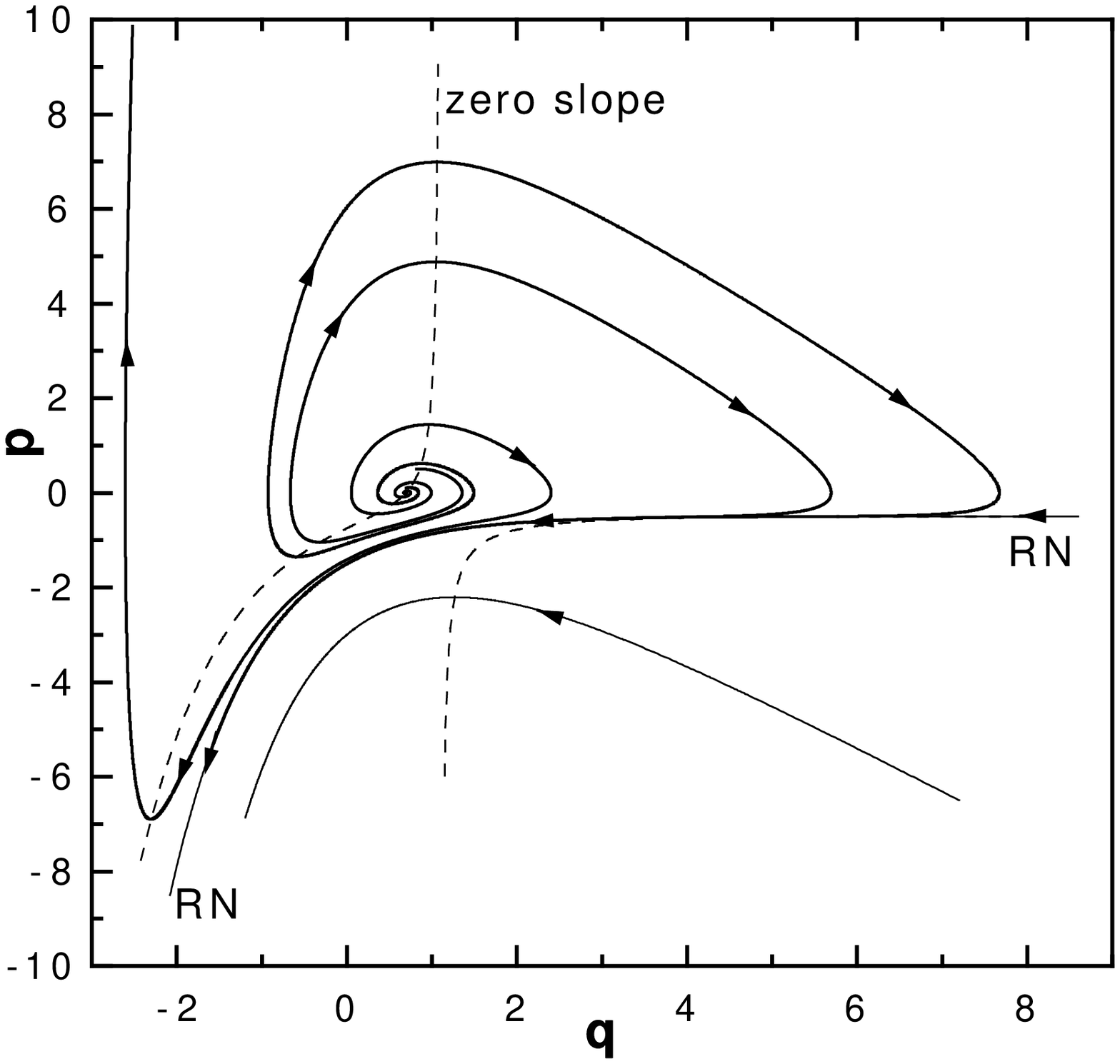}}
 \vspace{5mm}
 \caption{Phase portrait of the dynamical system (3.16), 
  {\sf RN}~--- an invariant set,
  corresponding to the RN--type solution (dashed~--- zero slope lines).}
 \label{phaseportrait}
\end{figure}

 Apart from the discrete picture given above one can find a truncated
 two-dimensional dynamical system clearly showing an infinitely
 oscillating nature of the generic solution. As we have already noted,
 the oscillation region starts with an exponential fall of $\Delta$
 which typically occurs after passing a local maximum $r^{max}$
 (Fig.\ \ref{oscilns}), and the right hand side of
 (\ref{eq4}) becomes comparatively small with respect to other
 terms. Another important feature is that the YM function $W$ (contrary to
 $\Delta$)  possesses a {\it finite\/} limit  $W_0$ in the singularity. 
 Omitting the right hand side of (\ref{eq4}), replacing $W$ by its 
 limiting value $W_0$, and neglecting $1$ as compared with $V^2/r^2$ 
 one can derive from the system (\ref{eq4}--\ref{eq5}) the following 
 two-dimensional dynamical system
\begin{eqnarray} \label{dsys}
      &&{\dot q} = p \, , \nonumber \\
      &&{\dot p} = (3e^{-q} - 1)p + 2 e^{- 2 q} - 1/2 \, ,
\end{eqnarray}
 where $\Delta = -(V_0^2/2) \exp(q)$,
 and a dot stands for derivatives with respect to $\tau = 2 \ln (r_h/r)$.
 This system has one (focal) fixed
 point ($p = 0$, $q = \ln 2$), corresponding to some imaginary charge
 RN-like local solution of the type (\ref{new}), 
 with eigenvalues $\lambda =(1 \pm i \sqrt{15})/4$. Its
 phase portrait is shown in Fig.\ \ref{phaseportrait}
 together with an invariant set
 $p = - e^{-q} - 1/2$ corresponding to the RN--type local solution 
 (\ref{rn}). The generic oscillating solutions lie above this curve.
 The phase motion in this region is unbounded, and there are
 no limiting circles. One can easily show that the rotation in the phase
 plane never stops and the limit $q = -\infty$ ($\Delta = 0$) can not
 be reached. The metric function $\Delta$ remains negative valued
 as $r \rightarrow 0$ and passes an infinite sequence of local maxima
 and minima. The class of oscillating solutions is stable 
 in a sense that a small deviation from one such solution moves us to another
 oscillating solution. 
 Thus the existence of the consistent two-dimensional truncation of 
 the full set of equations proves that the oscillating approach 
 to the singularity is generic for the EYM black holes. 
 The dynamical system (\ref{dsys}) was derived
 in our paper \cite{dgz} and reproduced in a slightly different notation
 in \cite{blm}.

%%%%%%%%%%%%%%%%%%%%%%%%%%%%%%%%%%%%%%%%%%%%%%%%%%%%%%%%%%%%%%%%%%%%%%%%%%%%%%

%%%%%%%%%%%%%%%%%%%%%%%%%%%%%%%%%%%%%%%%%%%%%%%%%%%%%%%%%%%%%%%%%%%%%%%%%%%%%%

\section{Power-law ``mass-inflationary'' singularity}
\renewcommand{\theequation}{4.\arabic{equation}}
\setcounter{equation}{0} 
In non-Abelian theories including scalar fields, such as EYMH and EYMD, 
another regime of the generic approach to the singularity is realized. 
In this case the scalar kinetic term becomes dominant soon after entering 
an asymptotic regime what results in a monotonous behavior of the
metric. Consider first the $SU(2)$ EYMH theory 
\begin{equation}\label{eymh}
     S = \frac{1}{16\pi} \, \int \,  \left \{- R - F^2
         + 2|D \Phi|^2 - \frac{\lambda}{2} (|\Phi|^2-\eta^2)^2 \right \}
         \sqrt{-g}\,d^4x \, ,
\end{equation}
\noindent
where  $\Phi$ is a Higgs field in
either vector (real triplet) or fundamental (complex doublet)
representations, $D \Phi$ is the corresponding YM covariant
derivative (in the doublet case $|D \Phi|^2=(D \Phi)^\dagger (D \Phi),\,
|\Phi|^2=\Phi^\dagger \Phi$) and without loss of generality both 
the Planck mass and the gauge coupling constant are set to unity. 
In the flat space-time the triplet version of the theory gives rise 
to regular magnetic monopoles, the doublet version~--- to sphalerons.
New physically interesting configurations emerge when gravity is coupled 
in a self-consistent way, in particular, static spherical
black holes exist in both cases: monopole  \cite{lnw} and
sphaleron \cite{gmn}. 

Static spherically symmetric configurations of the YM fields
are still given by the ansatz (\ref{eq3}), while the Higgs field is      
$\Phi^a T_a =\phi(r) \, T_r$  in the triplet case,
and $\Phi =\phi(r) v$ in the doublet one, where $v$ is
some (here irrelevant) spinor depending only on the angle variables. In both
cases $\phi(r)$ is the only real scalar function of the radial variable.

The system of equations following from (\ref{eymh}) with this ansatz
may be presented as a set of three coupled equations for
$W$, $\phi$, and $\Delta=r^2 - 2 m r$:
\begin{eqnarray}\label{h1}
 \left(\frac{\Delta}{r^2}W' \right)'+\frac{\Delta}{r}W' \phi'^2&=&
 \frac{1}{2}\frac{\partial {\cal V}}{\partial W} -Q \frac{W'}{r} \, , \\
 \left(\frac{\Delta}{r}\right)'+\Delta \phi'^2&=& 1-2{\cal V}-Q \, ,   \\
 \left(\Delta \phi' \right)'+\Delta r \phi'^3&=&
 \frac{\partial {\cal V}}{\partial \phi} -Q r \phi' \, , \label{h3}
\end{eqnarray}
where $Q = 2 \Delta W'^2/r^2$, and
\begin{equation}
{\cal V}\equiv {\cal V}(W,\phi,r)=\frac{V^2}{2 r^2}+
\frac{\lambda r^2}{8}(\phi^2-\eta^2)^2 + P^2,
\end{equation} 
with $P=W \phi$ in the triplet case and $P=(W+1) \phi$ in the doublet 
one. An equation for  $\sigma$ now becomes
\begin{equation}
(\ln \sigma)'=\frac{2}{r}W'^2 +r \phi'^2, 
\end{equation} 
and can be easily integrated once $W$, $\phi$ are found.

Local expansions of three types listed in the Sec.\ 2 can be easily
generalized to include Higgs \cite{blm}, and the same arguments
can be used to prove that neither of the corresponding global solutions
is generic. Meanwhile, the forth local branch can be found analytically
in this case \cite{gd}. It can be derived using the  
consistent truncation of the non-linear systems of equations.
An appropriate truncation consists in omitting from the
equations all matter terms except for those related to the gradient
of the scalar field. This reduces to dropping the right hand sides
in Eqs.\ (\ref{h1}--\ref{h3}).  
The resulting system can be easily disentangled leading to the following 
decoupled equations:
\begin{eqnarray}
     &&W''-\frac{W'}{r}=0 \, , \\
     &&\phi''+\frac{\phi'}{r}=0 \, ,
\end{eqnarray}
which can be solved as follows:
\begin{eqnarray}
     && W=W_0+ b r^2 \, , \\
     && \phi=\phi_0+ k \ln r \, .
\end{eqnarray}
Here $W_0$, $\phi_0$, $b$, $k$ are free parameters.
Note that contrary to the expansions discussed in the Sec.~2, this is not a
power series solution. Higgs field is logarithmically divergent, so that
its derivative diverges as $r^{-1}$, this is why the corresponding terms
become dominant for sufficiently small $r$.
Once $W$ and $\phi$ are found, the metric function $\Delta$ can be obtained 
by integrating the simple equation
\begin{equation}
     \frac{\Delta'}{\Delta}=\frac{1}{r} - r \phi'^2 \, ,
\end{equation}
what gives
\begin{equation}
     \Delta=-2 m_0 r^{(1-k^2)}
\end{equation} 
with the fifth (positive) constant $m_0$.
Hence, by counting free parameters, this is a generic
solution with non-positive $\Delta$. Now, to find whether
our truncation, one  has to substitute the solution into the Eqs.\ 
(\ref{h1}--\ref{h3})
and to check whether the right hand side terms are indeed comparatively 
small. One finds the following condition:  $k^2>1$.
This means that the metric function $\Delta$ is divergent at the singularity.
The corresponding mass-function is also divergent according
to the power-law
\begin{equation}
     m=\frac{m_0}{r^{k^2}} \, .
\end{equation}
The asymptotic behavior of $\sigma$ dominated by the scalar term then reads
\begin{equation}
     \sigma=\sigma_0 r^{k^2} \, ,
\end{equation}
with (positive) constant $\sigma_0$.

This local solution  can be interpreted as exhibiting a 
``power-law mass inflation''. As a matter of fact this regime
has nothing to do with the usual (exponential) mass-inflation 
which presumably takes place once a Cauchy horizon is approached.

The second example of the scalar-dominated singularity is given
by the EYMD theory:
\begin{equation}
                S = \frac{1}{16\pi} \int \left \{-R+ 2 (\nabla \phi)^2 - 
                    {\rm e}^{-2 \phi} F^2 \right \}\sqrt{-g} d^4x \, ,
\end{equation}
The equations of motion for $W$, $\Delta$, $\phi$ now take the form
\begin{eqnarray}
     && \Delta U'-2\Delta U \phi' = W V/r-{\cal F}W' \, , \label{eq3d} \\
     && (\Delta/r)'+\Delta \phi'^2={\cal F}-2 \Delta U^2 {\rm e}^{-2 \phi} \, ,
        \label{eq4d} \\
     && (\Delta \phi')' + \Delta r \phi'^3 =
        {\cal F} - 2 \Delta(\phi' r + 1) U^2 {\rm e}^{-2 \phi}-1 \, , 
        \label{eq5d}
\end{eqnarray}
 where
        ${\cal F} = 1 - V^2 {\rm e}^{-2 \phi}r^{-2} \, ,
        \quad V = W^2-1 \,.$
The remaining equation for $\sigma$ reads
\begin{equation} \label{eq6d}
     (\ln \sigma)' =r \left( \phi'^2 + 2 U^2 {\rm e}^{-2 \phi} \right)\,.
\end{equation}

 Similarly to the EYMH case for sufficiently 
 small $r$ the right sides of the Eqs.\ (\ref{eq3d})--(\ref{eq5d})
 become small in comparison with the left hand side terms and one
 gets the following truncated system
\begin{eqnarray}
     && \left(\ln U \right)'-2 \phi' = 0 \, , \nonumber \\
     && \left[ \ln (\Delta/r) \right]' = \left[ \ln (\Delta \phi')
        \right]' = -r \phi'^2 \,. \label{eq16}
\end{eqnarray}
 Its integration gives the following five-parameter (i.e., generic) 
 family of solutions  
\begin{eqnarray}
     && W = W_0 + b r^{2(1 - \lambda)} \, , \nonumber \\
     && \Delta = -2 \mu r^{(1 - \lambda^2)} \, , \label{eq17} \\
     && \phi = c + \ln \left( r^{-\lambda} \right) \, , \nonumber
\end{eqnarray}
 with constant  $W_0$, $b$, $c$, $\mu$, $\lambda$.
 The validity of the truncated equations (\ref{eq16}) now can be checked
 by substituting the asymptotic solution (\ref{eq17}) into the full system
 (\ref{eq3d})--(\ref{eq5d}). For consistency it is sufficient that the
 following inequalities hold:
\begin{equation}\label{pi}
     \sqrt{2} - 1 < \lambda < 1 \, ,
\end{equation}
 which is in agreement with the numerical data.

 From (\ref{eq17}) it follows that the mass function diverges as
 $r \rightarrow 0$ according to the power law:
\begin{equation}
     m(r) = \frac{\mu}{r^{\lambda^2}} \,.
\end{equation}
 The corresponding $\sigma$  tends to zero as
\begin{equation}
     \sigma(r)= \sigma_1 r^{\lambda^2} \, ,
\end{equation}
 where $\sigma_1 = \rm{const}$. A typical EYMD solution is shown 
 in Fig.\ \ref{firstoscln}.

 The only difference with the EYMH case is that the region of the
 power index (\ref{pi}) is different. This leads to different
 picture of the singularity in the Kantowski-Sachs interpretation:
 a point like singularity in the EYMH case, but a cigar singularity
 in the EYMD case \cite{gd}.

\section{Discussion}
\renewcommand{\theequation}{5.\arabic{equation}}
Non-Abelian black holes turned out to be a useful laboratory
to explore the nature of the singularity inside ``realistic'' black holes.
Unlike the  gravity coupled Abelian
field models such as Maxwell, Maxwell--dilaton--axion,
or more general string-inspired systems,
non-Abelian models have an advantage to give rise to several 
qualitatively different possibilities as far as the singularity is
concerned. Moreover, in spite of the absence of exact
analytic solutions, one can find a correspondence between
the singularity structure and external parameters such as the black 
hole mass. Thus the relative weight of different type interior structures
in the parameter space may be found. This opens a new way to probe
the Strong Cosmic Censorship using the genericity argument. An advantage
of this approach is that a non-trivial information may be extracted
already at the level of static (``eternal'') black hole solutions. 
One can speculate that non-Abelian electroweak theory should
replace the Maxwell electrodynamics at  high energies 
which are reached in the course of the
mass-inflation inside  a (perturbed) Reissner--Nordstr\"om black hole
due to the phase transition similar to that in cosmology. Therefore a
further fate of the black hole should be in the scope of a non-Abelian 
theory. Although the full time-dependent picture is likely to
be quite complicated, it is reassuring to feel that the
static spherical non-Abelian black holes choose the 
spacelike singularity in conformity with the Strong Cosmic Censorship.

The new features observed in the interiors of non-Abelian  black holes
are due to the field components which are not connected with the
conserved charges and are usually termed as a hair.
Violation of the no-hair conjecture  in black holes
traditionally was attributed to their external appearance. New
results clearly show that the hair is equally important
for the {\it internal\/} structure of black holes. Hair ``roots'' penetrate up 
to the singularity supplying it with additional degrees of freedom. A tiny
black hole inside a large magnetic monopole generically has
a very different internal structure than the Schwarzschild black hole.
Hairy black holes have hairy singularities.   

Qualitatively speaking, the interiors
of the EYM  black holes look like the
hair-perturbed Schwarzschild or Reissner--Nordstr\"om interiors.
So it is not surprising that one encounters an exponential growth
of mass whenever the metric reaches an ``almost'' Cauchy horizon.
This phenomenon, first noted in \cite{dgz} and termed as ``mass inflation''
in \cite{blm}, is, however, only  a half
of the story. In the oscillating regime one observes two qualitatively
different ``mass-inflations'': the first  is the local inflation at some
very short interval of each oscillation cycle, the second is 
associated with the exponential growth  of mass from cycle to cycle 
\cite{gdz} while the singularity is approached.

If scalar fields are present, the dynamics gets a new dimension. 
Once the scalar component is excited, it soon
becomes a dominant factor of the evolution which changes drastically
the approach to the singularity. In the scalar-dominated regime
no mass-inflation is manifest, instead one observes a behavior which we  
call a ``power-law mass-inflation'', i.e., a power-fashion
divergence of the mass-function near the singularity. This behavior
has a different origin as compared with the usual mass-inflation, in 
particular, it is not related to an approach to the Cauchy horizon. Rather
it follows from the coupled Einstein--scalar field dynamics which
is essentially Abelian. Indeed, the same behavior near the singularity
was earlier observed in the Kantowski--Sachs cosmology with a non-linear
(one-component) scalar field source. Still the non-Abelian nature of 
the model is substantial, otherwise the (asymptotically flat) static 
black holes are prohibited by the no-hair theorems.

 The analysis given here is purely classical, a few words are in order
 about the relevance of the results to the full quantum theory. 
 Vacuum polarization of the conformal scalar field 
 on the HMI background was considered in \cite{abc}. It was found 
 that the correction to the mass function diverges more strongly
 than in the classical case. Hence the singularity is not smoothened but 
 rather intensified. In the oscillating regime there is no hope
 to compute quantum effects quasi-classically. Moreover, huge values
 of the mass function (in Planck's units) encountered soon after entering 
 such a regime indicate that the quantum behavior of the model
 should be considered nonperturbatively and may well be 
 qualitatively different from the classical picture. However, the
 conclusion about the spacelike nature of the generic singularity 
 is unlikely to be changed.

 D.V.G. thanks the Organizing Committee for hospitality and support
 during the Workshop. The research was supported in part by the 
 RFBR grants 96--02--18899, 18126.


\begin{thebibliography}{99}
\bibitem{rw}
R.~Ruffini and J.~A.~Wheeler, Physics Today {\bf 24}, 30 (1970).

\bibitem{bk} 
R.~Bartnik, J.~McKinnon, Phys. Rev. Lett., {\bf 61}, 141 (1988).

\bibitem{sph} 
D.~V.~Gal'tsov, M.~S.~Volkov, Phys. Lett., {\bf 273}, 255 (1991).

\bibitem{str}
N.~Straumann, Z.~H.~Zhou, Phys. Lett., {\bf 237}, 353 (1990). 

\bibitem{sw}
D.~Sudarsky and R.~M.~Wald, Phys. Rev. {\bf D 46}, 1453 (1992).

\bibitem{snm}
M.~S.~Volkov, D.~V.~Gal'tsov, Phys. Lett., {\bf 341}, 279 (1995).

\bibitem{mw}
I.~Moss and A.~Wray, Phys. Rev. {\bf D 46}, 1215 (1992).

\bibitem{vgkb} 
 M.~S.~Volkov and D.~V.~Gal'tsov, Pis'ma Zh.\  Eksp.\  Teor.\  Fiz.,
 {\bf 50}, 312 (1989) [JETP Lett., {\bf 50}, 345 (1990)];
 H.~P.~Kunzle, A.~K.~M.~Masood--ul--Alam, J.\  Math.\  Phys.\ 
 {\bf 31}, 928 (1990);
 P.~Bizon, Phys.\  Rev.\  Lett., {\bf 64}, 2844 (1990).

\bibitem{vg} 
M.~S.~Volkov and D.~V.~Gal'tsov, Sov.\  J.\  Nucl.\  Phys., {\bf 51},
747 (1990).

\bibitem{dil} 
D.~V.~Gal'tsov and E.~E.~Donets, 
Int.\ J.\ Mod.\ Phys.\ {\bf D 3}, 755 (1994) and refences therein.

\bibitem{lnw}
K.~Lee, V.~P.~Nair, and E.~Weinberg, 
Phys. Rev. {\bf D 45}, 2751 (1992);
M.~Ortiz, Phys. Rev. {\bf D 45}, 2586 (1992);
P.~Breitenlohner, P.~Forg\'acs, and D.~Maison, Nucl. Phys.
{\bf B 383}, 357 (1992); {\bf B 442}, 126 (1995);
T.~Tachizawa, K.~Maeda, and T.~Torii, Phys. Rev. {\bf D 51}, 4051 (1995).

\bibitem{gmn} 
B.~R.~Greene, S.~D.~Mathur, and C.~M.~O'Neill,
Phys. Rev. {\bf D 47}, 2242 (1993).

\bibitem{tm}
T.~Torii, K.~Maeda, Phys. Rev. {\bf D 48}, 1643 (1993).

\bibitem{dm}
D.~Maison, {\em Solitons of the Einstein--Yang--Mills Theory}, 
gr-qc/9605053.

\bibitem{dgz}
E.~E.~Donets, D.~V.~Gal'tsov, and M.~Yu.~Zotov, 
{\it Internal structure of Einstein--Yang--Mills black holes},
gr-qc/9612067, Phys.\ Rev.\ {\bf D 56}, 3459-3465 (1997).

\bibitem{bkl}
V.~A.~Belinskii, I.~M.~Khalatnikov, and E.~M.~Lifshitz,
Adv.\ Phys. {\bf 19}, 525 (1970).

\bibitem{blm}
P.~Breitenlohner, G.~Lavrelashvili, and D.~ Maison, 
{\it Mass inflation and chaotic behavior inside hairy black holes},
MPI-PhT/97-20, BUTP-97/08, gr-qc/9703047.

\bibitem{gd}
 D.~V.~Gal'tsov and E.~E.~Donets,
 {\it Power--law mass inflation in Einstein--Yang--Mills--Higgs black holes}
 gr-qc/9706067.

\bibitem{gdz}
 D.~V.~Gal'tsov, E.~E.~Donets, and M.~Yu.~Zotov,
 Pis'ma Zh.\  Eksp.\  Teor.\  Fiz.,
 {\bf 65}, 855 (1997), [JETP Lett., {\bf 65}, 895 (1997)]; (gr-qc/9706063).

\bibitem{minfl} 
E.~Poisson and W.~Israel, Phys.\ Rev.\ Lett. {\bf 63}, 1663 (1989);
Phys.\ Rev.\ {\bf D 41}, 1976 (1990);
A.~Ori, Phys.\ Rev.\ Lett. {\bf 67}, 789 (1991);
A.~Bonnano, S.~Droz, W.~Israel, S.~M.~Morsink, 
Phys.\ Rev.\ {\bf D 50}, 7372 (1994); 
A.~Ori, Phys.\ Rev.\ {\bf 55}, 3575 (1997). 
 
\bibitem{pdm}
B.~Paul, D.~P.~Datta, and S.~Mukherjee,
Modern Physics Letters {\bf A1}, No.~2, 149 (1986).

\bibitem{blm2}
P.~Breitenlohner, G.~Lavrelashvili, and D.~Maison, 
{\it Non-Abelian black holes: The inside story},
BUTP-97/23, gr-qc/9708036.

\bibitem{pg} 
D.~N.~Page, in Proc.\ NATO Advanced Summer Institute on 
Black Hole Physics (Erice 1991), ed.\ V.~De~Sabbata and Z.~Zhang 
(Kluwer, 1992), p.\ 185.

\bibitem{abc} W.~G.~Anderson, P.~R.~Brady, and R.~Camporesi, 
Class.\ Quant.\ Grav. {\bf 10}, 497 (1993).

\bibitem{smw}
 J.~A.~Smoller and A.~G.~Wasserman {\it Reissner--Nordstr\"om--like
 solution of the SU(2) Einstein--Yang/Mills equations}, gr-qc/9703062.

\end{thebibliography}
\end{document}